\documentclass[journal]{IEEEtran}

\IEEEoverridecommandlockouts
\normalsize
\usepackage{amssymb}
\usepackage{verbatim}
\usepackage{bbold}
\usepackage{bbm}
\usepackage[cmex10]{amsmath}
\usepackage{stfloats}
\usepackage{multirow}
\usepackage{tabularx}
\usepackage{booktabs}
\usepackage{amsmath}
\usepackage{epsfig,epsf,color,balance,cite}
\usepackage{indentfirst}
\usepackage{mathrsfs}
\usepackage{float}
\usepackage{algorithm}
\usepackage{algorithmic}
\usepackage{graphicx}  
\usepackage{subcaption}

\usepackage{dsfont}
\usepackage{color}
\usepackage{url}
\usepackage{color}
\usepackage{graphicx}
\usepackage{epstopdf}
\usepackage{enumerate}
\usepackage{cite}

\begin{document}
\title{NTP-INT: Network Traffic Prediction-Driven In-band Network Telemetry for High-load Switches
\\
}

\author{
\IEEEauthorblockN{ Penghui Zhang,~\IEEEmembership {Student Member,~IEEE,} Hua Zhang,~\IEEEmembership {Member,~IEEE,}  
Yuqi Dai,~\IEEEmembership {Student Member,~IEEE,} Cheng Zeng,~\IEEEmembership {Student Member,~IEEE,} Jingyu Wang, ~\IEEEmembership {Member,~IEEE,} Jianxin Liao, ~\IEEEmembership {Member,~IEEE}
}\\
\thanks{Penghui Zhang, Hua Zhang, Yuqi Dai and Cheng Zeng are with the National Mobile Communications Research Laboratory, Southeast University, Nanjing 211111, 
China. (email: phzhang@seu.edu.cn, huazhang@seu.edu.cn, 230228198@seu.edu.cn, czeng@seu.edu.cn)}
\thanks{Jingyu Wang and Jianxin Liao are with the State Key Laboratory of Networking and Switching Technology, Beijing University of Posts and Telecommunications, Beijing 100876, China. (email: wangjingyu@bupt.edu.cn, liaojx@bupt.edu.cn)}
}

\maketitle
\vspace{-2.5em}\vspace{-2.5em}
\begin{abstract}
In-band network telemetry (INT) is essential to network management due to its real-time visibility. However, because of the rapid increase in network devices and services, it has become crucial to have targeted access to detailed network information in a dynamic network environment. 
This paper proposes an intelligent network telemetry system called NTP-INT to obtain more fine-grained network information on high-load switches. Specifically, NTP-INT consists of three modules: network traffic prediction module, network pruning module, and probe path planning module.
Firstly, the network traffic prediction module adopts a Multi-Temporal Graph Neural Network (MTGNN) to predict future network traffic and identify high-load switches. Then, we design the network pruning algorithm to generate a subnetwork covering all high-load switches to reduce the complexity of probe path planning. Finally, the probe path planning module uses an attention-mechanism-based deep reinforcement learning (DEL) model to plan efficient probe paths in the network slice.
The experimental results demonstrate that NTP-INT can acquire more precise network information on high-load switches while decreasing the control overhead by 50\%.

\end{abstract}
\begin{IEEEkeywords}

Network telemetry, Network measurement, Network traffic prediction, Graph neural network, Deep reinforcement learning

\end{IEEEkeywords}

\footnotetext{This work was supported by the National Key Research and Development Program of China under Grant(2020YFB1807803) 
.}

\IEEEpeerreviewmaketitle
\vspace{-1em}
\section{Introduction}
\label{section1}
In-band network telemetry (INT) plays an essential role in real-time network management in modern communication systems, including mobile radio, vehicular communication, and cellular networks \cite{1}. By allowing switches to add information to the passing packets, INT enables flexible and real-time network monitoring \cite{5}. This capability is crucial for applications such as congestion control, load balancing, fault location, and the monitoring of mobile radio systems, including vehicular communications\cite{2, 4}.  
As 6G technologies continue to evolve, the need for efficient management of dynamic and high-traffic environments like vehicular networks becomes even more important.

With the rapid development of 6G networks, new technologies like network function virtualization (NFV) and service function chaining (SFC) are expected to significantly increase the complexity of network services.   These advancements bring about greater loads on critical infrastructure, such as high-load switches \cite{8}, which are responsible for handling substantial traffic in real-time. 
High-load switches have a higher probability of experiencing network congestion and failures and are critical to ensuring overall network performance.
Therefore, for more effective and timely network management, it becomes imperative for INT to acquire more fine-grained network information, specifically from high-load switches. 

Because of the uncontrolled probe paths, traditional INT makes it difficult to solve the problem of more fine-grained telemetry with specific switches. To obtain more fine-grained network information, the common approach is to increase the telemetry frequency of the entire network \cite{9}. However, this simple approach comes with significant overhead, consuming vast amounts of additional bandwidth resources.
Moreover, due to the lack of a targeted probe planning strategy \cite{10,11}, network resources often fail to be accurately allocated to high-load switches that need to be probed at high frequencies.
In contrast, due to controllable probe paths \cite{12}, active network telemetry (ANT) can develop a targeted high-frequency probe path planning strategy based on the position information of the high-load switches, thereby significantly improving the utilization efficiency of network resources \cite{8,13}. However, to obtain network information more effectively, ANT still faces the following two challenges:
\begin{itemize}
\item How to identify high-load switches in a dynamic network environment \cite{14}. Identifying high-load switches in a dynamic network environment is challenging due to constantly changing network traffic \cite{16}, especially in 6G networks.  The telemetry strategies \cite{15} designed according to the current network conditions may no longer be applicable at the time of execution, thus affecting the real-time and accuracy of the telemetry system.
\item How to efficiently plan high-frequency probe paths \cite{1}. For large networks, the complexity of probe path planning over the whole network is very high. The existing path planning algorithm cannot give the optimal solution quickly \cite{21, 20}. Moreover, to reduce the telemetry overhead, it is necessary to avoid excessive telemetry in non-critical network areas \cite{25}.
\end{itemize}

Unfortunately, existing ANT still faces significant challenges in 6G networks (discussed in Section \ref{section2}). Since the network environment is dynamic and bursts of traffic change very rapidly, it is difficult for them to accurately identify high-load switches, where network traffic prediction techniques can overcome this challenge well \cite{17}. 
Typical network traffic prediction techniques include Support Vector Machines (SVM) \cite{36}, Long Short-Term Memory Networks (LSTM) \cite{37}, Graph Neural Networks (GNN) \cite{38},  etc., which provide new directions for telemetry systems identify high-load switches that need to collect their information at a higher frequency \cite{18, 19}.
To overcome the second challenge, Deep Reinforcement Learning (DRL) \cite{22} provides a new direction for efficient planning of high-frequency probe paths \cite{NetworkAI}. Currently, there is research applying DRL to network telemetry, known as AdapINT \cite{AdapINT}. DRL offers more flexible solutions for probe path planning. However, as the network scale grows, the training time of the DRL model increases exponentially due to the vast action space.
To reduce the training time of the DRL model, an independent functional virtual subnetwork, which is partitioned from the complex network structure and covers all high-load switches\cite{24, 26}, can significantly reduce the action space of the DRL model, avoiding excessive telemetry in less critical network areas\cite{25}.

This paper proposes an intelligent telemetry system called NTP-INT, which can obtain more fine-grained network information from high-load switches. The NTP-INT comprises three main components: a network traffic prediction module, a network pruning module, and a probe path planning module. 
Specifically, the network traffic prediction module combines INT with a Multi-Temporal Graph Neural network (MTGNN) model to predict network future traffic and identify high-load switches based on the prediction results \cite{27}. 
The network pruning module focuses on generating a subnetwork that covers all high-load switches \cite{28}, and we design the network pruning algorithm, which comprises several steps: generating sub-connected graphs, detecting connection points, and creating biconnected graphs. 
The task of the probe path planning module is to plan the paths of the high-frequency probes. 
We use a DRL model based on an attention mechanism to design a high frequency probe path planning scheme and design a mask function to speed up the training of the model.
The simulation results show that the network traffic prediction module can capture the characteristics of network traffic more accurately. The network pruning module effectively reduces probe path planning complexity and halves the DRL training time. Our DRL-based probe path planning significantly reduces control overhead by 50\% without compromising accuracy.

The main contributions of this paper are as follows:

\vspace{-0em}
\begin{itemize}
\item We propose NTP-INT, an intelligent network telemetry system for high-load switches, targeted to obtain fine-grained network information on high-load switches and minimize the telemetry overhead.
\item To predict sudden traffic changes, we integrate the MTGNN model with the INT system and optimize it to enhance prediction accuracy.  This enables the network telemetry system to proactively anticipate and respond to traffic fluctuations in real-time.
\item To reduce the training time of the DRL model and avoid excessive telemetry in unimportant network areas, we design the network pruning algorithm to generate a subnetwork covering all high-load switches. The produced subnetwork reduces the action space of DRL and ensures the backup path.
\item To reduce the control overhead, we use a DRL model based on an attention mechanism to plan the probe paths. Considering the characteristics of the probe path planning problem, we design a mask function to speed up model training.
\end{itemize}

The rest of this paper is organized as follows:  Section \ref{section2} describes the related work. Section \ref{section3} introduces the architecture of NTP-INT. Section \ref{section4}-\ref{section6} respectively describe the network traffic prediction module, network pruning module, and probe path planning module. Section \ref{section6} presents the evaluation results. Finally, Section \ref{section7} concludes this paper.

\section{Related Work}
\label{section2}
This section provides a comprehensive review and analysis of related work, including network telemetry and traffic prediction, which will serve as a foundation for future research.

\subsection{Network Telemetry}
In recent years, the development of Software-Defined Networking (SDN) and Programmable Data Planes (PDP) has significantly enhanced network telemetry capabilities. INT, by embedding real-time network status information into data packets, provides fine-grained, low-latency network monitoring, enabling more accurate network performance management. However, traditional INT methods are often limited in their ability to adapt to rapid network topology changes and high-frequency telemetry requirements. As network complexity increases, particularly with the introduction of high-load switches and dynamic traffic patterns, traditional INT systems struggle to offer both high precision and efficiency \cite{5}. INT uses data packets within the network to collect network information, which can be further classified into passive network telemetry (PNT) and ANT.

PNT does not actively inject probes into the network. Instead, it relies on existing packets within the network to collect network information. Compared to traditional network measurement techniques, 
Typical PNT systems, such as Sel-INT \cite{33}, PINT \cite{10}, and INT-label \cite{11}, offer finer-grained measurements, real-time capabilities, flexibility, scalability, and data consistency. These systems provide network managers with precise and comprehensive network state information, enabling more effective network management and optimization.

However, a significant problem of PNT lies in its inability to obtain comprehensive network-wide information due to the uncertainty of existing packet forwarding paths. To solve this problem, ANT actively sends probes to collect network information along user-specified forwarding paths. Typical ANT systems, such as INT-path \cite{12}, IntOpt \cite{8}, and NetView \cite{13}, can achieve full network coverage and enhanced flexibility. 
Nevertheless, existing ANT systems still face challenges in balancing telemetry overhead and accuracy. High-precision network telemetry often leads to significant telemetry overhead, potentially impacting network performance \cite{9}. 
These network telemetry systems can not adapt well to dynamic network environments and diversified telemetry requirements.

To address this issue, AdapINT is proposed as a DRL-based in-network telemetry system \cite{AdapINT}. Facing dynamic network environments, AdapINT brings flexible and adaptive solutions for probe path planning \cite{20}. The strong generalization capability of DRL enables it to quickly adapt and make wise decisions in new or unknown network environments, ensuring the continuous effectiveness of probe path planning \cite{23}. However, DRL introduces other pressing issues to AdapINT. Firstly, in response to sudden traffic changes in the network, AdapINT cannot pre-adjust the probe paths. This significantly reduces its ability to cope with sudden traffic changes. Secondly, for large-scale networks, the action space in the DRL model becomes extremely large, which causes the model training time to grow exponentially and severely affects the system's operation.

\subsection{Network Traffic Prediction}

Network traffic prediction technology is important in a dynamic network environment. Due to the constant changes in network traffic \cite{16}, traditional telemetry strategies based on the current state of the network can quickly become obsolete, affecting the real-time and accuracy of data collection \cite{15}. Network traffic prediction technology can use historical data to predict future network traffic \cite{17}, accurately identify high-load switches, and optimize data collection frequency, significantly improving the efficiency of telemetry systems.

Traditional traffic prediction technologies primarily use statistical characteristics derived from historical data, such as the Autoregressive Integrated Moving Average (ARIMA) model \cite{35}. However, with the advent of machine learning, there has been a significant evolution in traffic prediction methods. These advancements include the utilization of SVM \cite{36}, LSTM \cite{37}, GNN \cite{38}, and so on. These modern methods have demonstrated remarkable proficiency in capturing intricate patterns and dynamics within traffic data, thereby enhancing prediction accuracy.

Integrating traffic prediction with network telemetry can mitigate the challenges posed by dynamic network environments. However, existing prediction methods typically rely on historical data and fail to adapt to rapid network changes, such as topology shifts or traffic surges \cite{40}. The key challenge is effectively integrating telemetry technology to capture more accurate and comprehensive traffic data \cite{39}. Moreover, selecting the right prediction model for different network scenarios remains an open issue \cite{27}.

Several studies have explored combining network telemetry with traffic prediction to enhance monitoring efficiency. For example, integrating real-time telemetry data from INT with machine learning techniques like SVM or LSTM has improved prediction accuracy. However, these methods still struggle to adapt to sudden network changes quickly. The Multi-Temporal Graph Neural Network (MTGNN) applied in this paper addresses these issues by leveraging both real-time INT data and temporal traffic dependencies, offering more accurate and timely predictions, especially in dynamic environments.

\section{System Design}
\label{section3}
In this section, we introduce the telemetry architecture of NTP-INT and subsequently detail its workflow, including the network traffic prediction module, network pruning module, and probe path planning module.

\begin{figure}
\centering
\setlength{\abovecaptionskip}{0.cm}

\includegraphics[width=\linewidth]{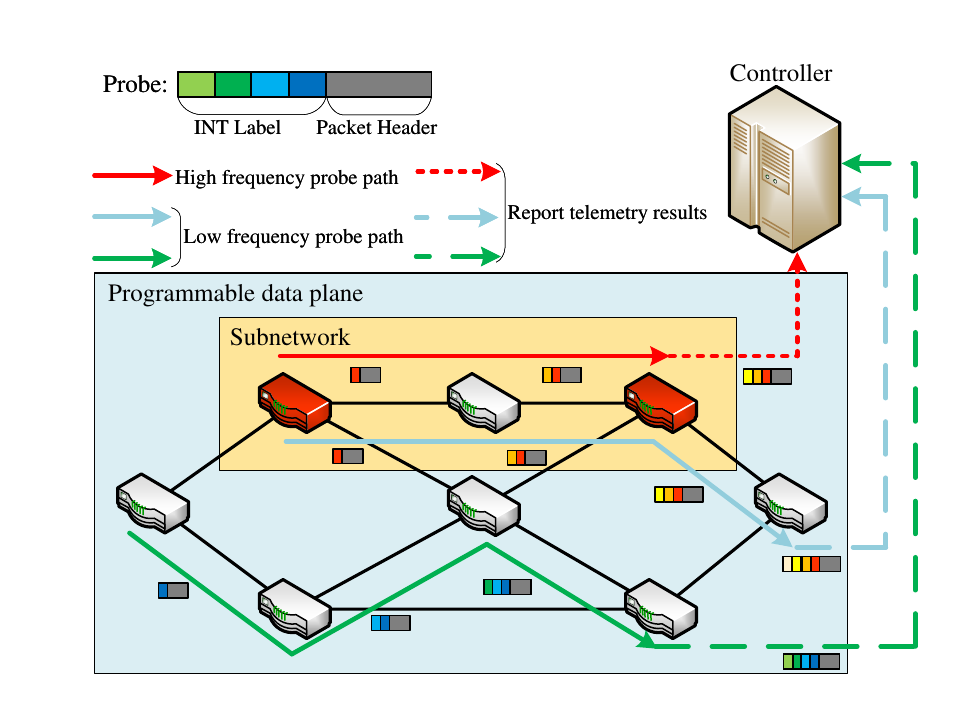}
\caption{Architecture of NTP-INT. }
\label{fig1}
\end{figure}
\subsection{Architecture of NTP-INT}
NTP-INT is committed to designing an efficient telemetry strategy for high-load switches and deploying additional high-frequency probes.

Fig.\ref{fig1} shows the architecture of NTP-INT. The light yellow area indicates the subnetwork. The green and blue probe paths represent low-frequency probe paths that cover all network devices, while the red probe paths specifically represent additional high-frequency probe paths deployed for high-load switches.

NTP-INT uses ANT's system architecture, where the controller dynamically updates telemetry policies based on the dynamic network environment \cite{12}. Next, we explain ANT's idea in detail. The probe is periodically injected into the programmable data plane. Due to the predefined probe format, the switch can recognize the probe and encapsulate the network information into metadata, which is then inserted into the probe. These probes are then forwarded along the user-defined path to the next port. When the probe reaches the final switch, the switch forwards it to the controller for analysis.
It is worth noting that to obtain finer-grained network information, NTP-INT deploys additional high-frequency probes for high-load switches, which are the design focus of NTP-INT and are described in detail in the next section.

\begin{figure}
\centering
\setlength{\abovecaptionskip}{0.cm}

\includegraphics[width=7cm]{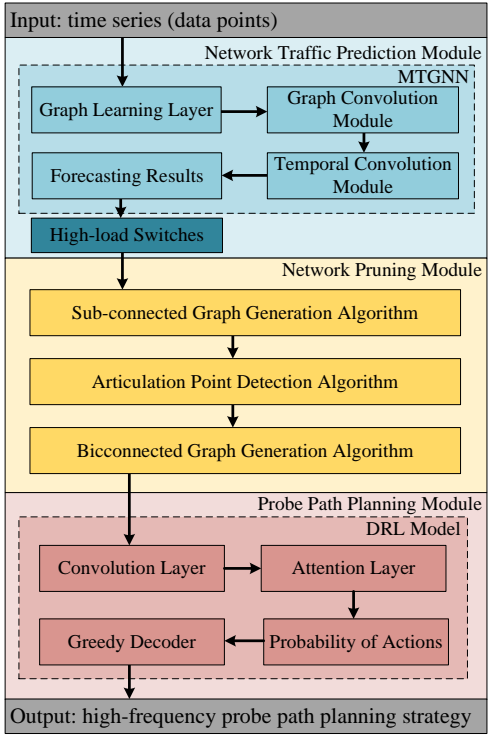}
\caption{Workflow of NTP-INT. }
\label{fig2}
\end{figure}
\vspace{-0.5em}
\subsection{NTP-INT Workflows for High-load Network Areas}

This subsection provides a comprehensive overview of NTP-INT's workflow for high-load switches, as depicted in Fig. \ref{fig2}. The workflow encompasses three crucial components: network traffic prediction module, network pruning module, and probe path planning module.

\emph{1) Network Traffic Prediction Module}:
The task of the network traffic prediction module is to predict future network traffic and identify high-load switches accordingly.
We use the MTGNN model to design the network traffic prediction model \cite{27} of NTP-INT. MTGNN combines the strengths of RNN and GNN to capture long-term dependencies in time series data and process data with complex spatial relationships. After the predicted results are obtained, high-load switches are identified, providing a basis for network pruning.

\emph{2) Network Pruning Module}:
To cover all high-load switches, the network slicing algorithm is designed, including the subconnected graph generation, articulation point detection, and biconnected graph generation. Considering the complexity of probe path planning and the stability of the telemetry system, the produced network slicing ensures the backup paths and the coverage area is as small as possible.

\emph{3) Probe Path Planning Module}:
The task of the probe path planning module is to complete the planning of a high-frequency probe. Thus, a self-learning DRL model based on an attention mechanism is designed. Considering the routing path planning problem's characteristics, we improve the RNN model's input queue and set the mask function to reduce the model complexity \cite{41}.

The detailed designs of the network traffic prediction module, network pruning module, and probe path planning module are discussed in detail in the following sections.

\vspace{-1em}
\section{Network Traffic Prediction Module}
\label{section4}
This section introduces the network traffic prediction module and explains the implementation process in detail.

\subsection{Module Overview}

We propose an enhanced MTGNN model integrated with INT to address the challenges of predicting network traffic in dynamic environments \cite{17}. Traditional models often rely on static topologies and historical data, which struggle to adapt to the rapid changes in modern networks, such as topology alterations and traffic bursts. INT, by providing real-time feedback on network conditions—such as traffic volume, link statuses, and topology changes—offers a more accurate and up-to-date view of the network, which is crucial for maintaining prediction accuracy.

To leverage the strengths of INT, we introduce several modifications to the MTGNN model \cite{38}. First, the input layer is dynamically adjusted to process real-time traffic data and topology updates from INT, ensuring the model always receives current network information. As network topologies frequently change, the graph structure in the improved MTGNN is updated incrementally, focusing only on the affected areas rather than recalculating the entire graph, thus enhancing computational efficiency. Additionally, the model integrates real-time feedback from INT, allowing it to compare predicted traffic with actual data, calculate prediction errors, and adjust its parameters through backpropagation, improving its accuracy over time. Finally, rather than retraining the entire model with each new batch of data, we implement incremental learning, allowing the model to continuously adapt to new data without requiring a full retraining cycle.

These modifications make MTGNN highly adaptable to dynamic network conditions, improving its ability to predict traffic accurately in real-time while handling the complexities introduced by ever-changing network topologies and real-time telemetry data.

\begin{figure*}
\centering
\setlength{\abovecaptionskip}{0.cm}

\includegraphics[width=16cm]{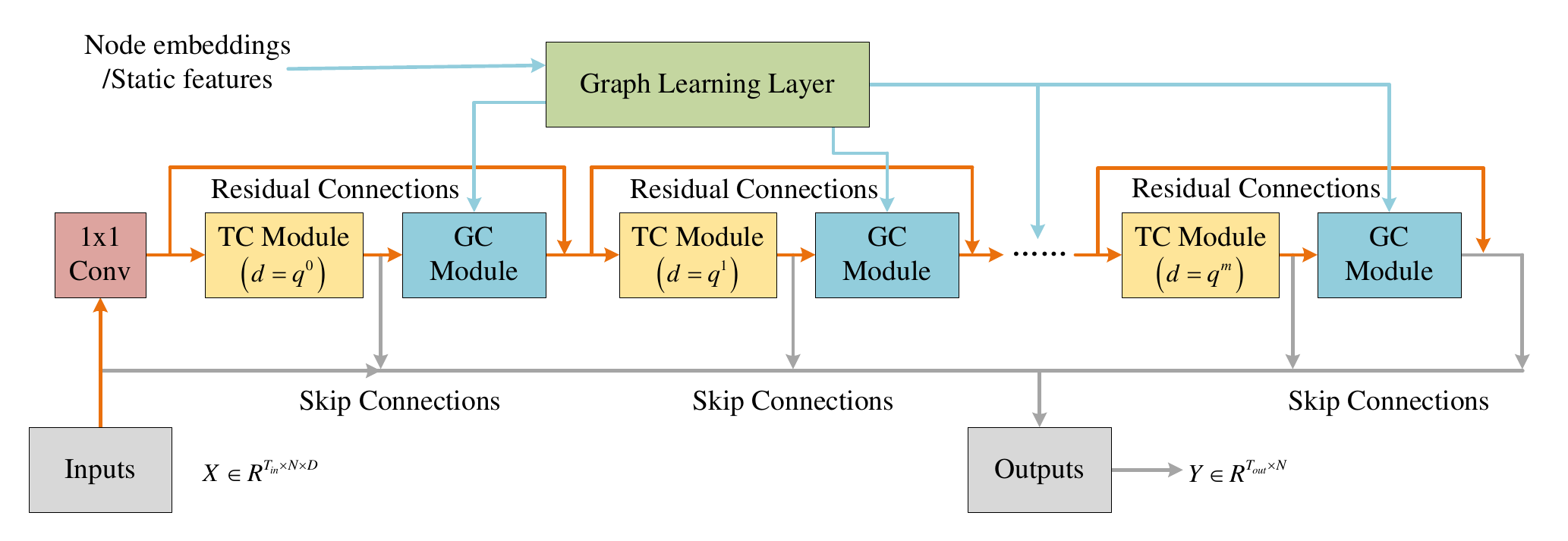}
\caption{The framework of the MTGNN. }
\label{fig3}
\vspace{-0.1cm}
\end{figure*}

\vspace{-0.5em}
\subsection{MTGNN Framework}

The MTGNN framework is comprised of three main components: the graph learning layer, the graph convolution module, and the temporal convolution modules \cite{27}. The graph learning layer aims to identify hidden associations between nodes by computing the graph adjacency matrix and using it as input for all graph convolution modules. Spatial and temporal dependencies are captured by interleaving graph convolution modules with temporal convolution modules. To avoid gradient vanishing, residual connections are added from the inputs of the time convolution module to the outputs of the graph convolution module. Lastly, skip connections are added after each temporal convolution module. The final output is generated by projecting the hidden features to the desired output dimension in the output module. Fig.\ref{fig3} provides an overview of the MTGNN framework, and the core components of MTGNN are illustrated in the following:

\emph{1) Graph Learning Layer}: The graph learning layer aims to learn an adaptive adjacency matrix that captures spatial relationships between variables in time series data.  It is important to note that this learned adjacency matrix is asymmetric.  The graph learning layer we employ is specifically designed to extract uni-directional relationships \cite{42}, illustrated as follows:

\vspace{-0.2cm}
\begin{IEEEeqnarray}{rCl} 
\label{form_1}
{\!}
\begin{split}
{{M}_{1}}=\tanh \left( \alpha {{E}_{1}}{{\Theta }_{1}} \right)\\
\end{split}
\end{IEEEeqnarray}

\vspace{-0.7cm}
\begin{IEEEeqnarray}{rCl} 
\label{form_2}
{\!}
\begin{split}
{{M}_{2}}=\tanh \left( \alpha {{E}_{2}}{{\Theta }_{2}} \right)\\
\end{split}
\end{IEEEeqnarray}

\vspace{-0.7cm}
\begin{IEEEeqnarray}{rCl} 
\label{form_3}
{\!}
\begin{split}
A=\text{ReLU}\left( \tanh \left( \alpha \left( {{M}_{1}}M_{2}^{T}-{{M}_{2}}M_{1}^{T} \right) \right) \right)\\
\end{split}
\end{IEEEeqnarray}

\vspace{-0.7cm}
\begin{IEEEeqnarray}{rCl} 
\label{form_4}
{\!}
\begin{split}
\text{idx}=\emph{argtopk}\left( A\left[ i,: \right] \right),i=1,2,\cdots ,N\\
\end{split}
\end{IEEEeqnarray}

\vspace{-0.7cm}
\begin{IEEEeqnarray}{rCl} 
\label{form_5}
{\!}
\begin{split}
A\left[ i,-\text{idx} \right]=0,i=1,2,\cdots ,N,\\
\end{split}
\end{IEEEeqnarray}
where ${{E} _ {1}} $, ${{E} _ {2}} $ represent randomly initialized node embeddings, ${{\Theta} _ {1}} $, ${{\Theta} _ {2}} $ represent the model parameters, $\alpha$ is a hyper-parameter for controlling the saturation rate of the activation function, and $\emph{argtopk}\left(\centerdot \right)$ returns the index of the top-k largest values of a vector. 
$A\in {{R}^{N\times N}}$ represents the adjacency matrix of a graph.
Eq. \ref{form_3} calculates the asymmetric information of the adjacency matrix $A$, where ReLU activation can regularize the effect of the adjacency matrix.
Meanwhile, Eq. \ref{form_4}-\ref{form_5} helps create a sparse adjacency matrix that reduces the computational cost of subsequent graph convolutional networks.  For each node, the top-k closest nodes are selected as its neighbors. While retaining the weights for connected nodes,  the weights of non-connected nodes are set to zero.

\begin{figure}[htbp] 
  \centering  
  \begin{minipage}{0.5\linewidth} 
    \centering  
    \includegraphics[width=.9\linewidth]{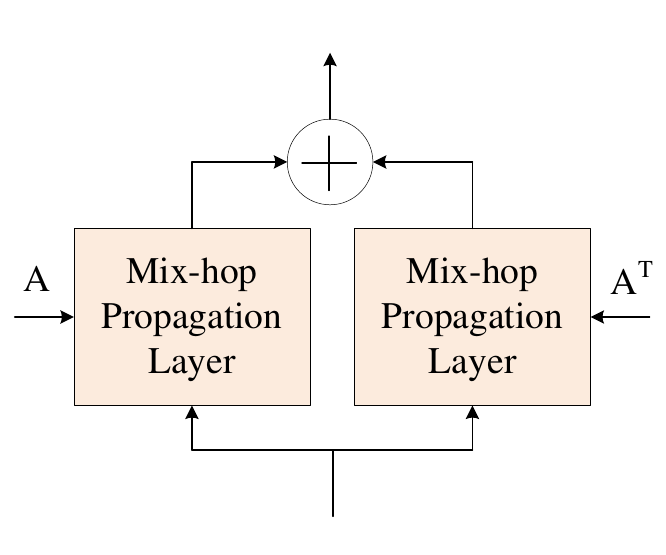} 
    \caption{Graph convolution module}  
    \label{fig4}  
  \end{minipage}%
  \begin{minipage}{0.5\linewidth} 
    \centering  
    \includegraphics[width=.9\linewidth]{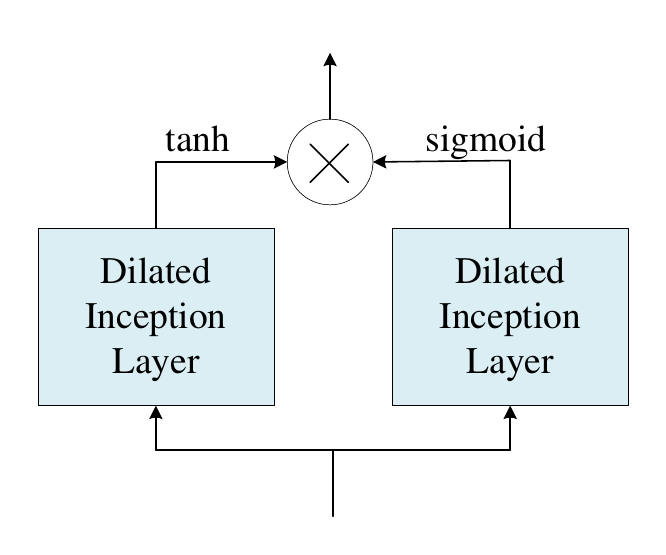} 
    \caption{Temporal convolution module}  
    \label{fig5}  
  \end{minipage}  
\end{figure} 
\emph{2) Graph Convolution Module}: The graph convolution module integrates both node and neighbor node information. As shown in Fig. \ref{fig4}, it consists of two mix-hop propagation layers, with horizontal and vertical movements corresponding to information propagation and information selection, respectively.

\emph{3) Temporal Convolution Module}: The temporal convolution module is responsible for extracting high-dimensional temporal features, and it achieves this by utilizing multiple standard one-dimensional expansive convolution kernels.  As depicted in Fig. \ref{fig5}, the temporal convolution module consists of two dilated inception layers.  The effectiveness of this structure has previously been validated in the field of computer vision.

\subsection{Implementation Steps}

This subsection details the implementation process, which comprises six key steps: Dynamic Input Layer Adjustment, Data Preprocessing, Incremental Graph Convolution, Real-time Feedback Integration, MTGNN Learning Algorithm, and High-load Switch Identification.

\emph{1) Dynamic Input Layer Adjustment}:
To effectively process real-time network data from INT, the model's input layer must be dynamically adapted to handle both network traffic data and topology information. Network traffic data is collected at specific time intervals and stored as a matrix, where each element represents the amount of traffic between a pair of nodes at a specific time. Topology data, which includes node connectivity and link statuses, is also updated in real-time from the INT system.

\emph{2) Data Preprocessing}: We preprocess the traffic data by removing outliers and normalizing the values to ensure that the model receives stable and accurate inputs. For each time step, the raw data is transformed into a format that can be directly fed into the MTGNN model, including time-series traffic data and real-time topology updates. This preprocessing step is essential to ensure the model's stability and accuracy during training and prediction.

\emph{3) Incremental Graph Convolution}: Due to frequent network topology changes, such as link failures or reconfigurations, the graph structure in MTGNN must be updated dynamically. To optimize performance, we use incremental graph convolution, which updates only the parts of the graph affected by topology changes rather than recalculating the entire graph. This method ensures that updates are performed efficiently by limiting computation to the relevant nodes and edges. The incremental approach helps maintain high efficiency while adapting to dynamic network changes.

\emph{4) Incremental Graph Convolution}: Real-time feedback from the INT system provides critical information about the network’s current state. This feedback, which includes data such as traffic volume, latency, and link status, is integrated with the model's predictions. The model compares its predicted traffic with real-time feedback and calculates the error, $error = y_{real}-y_{pred}$, which is then used to update the model’s parameters through backpropagation. The real-time feedback loop allows the model to adjust its predictions continuously, enhancing its accuracy and adaptability in dynamic environments. By incorporating real-time feedback, the model can learn to correct its predictions in near real-time, improving its predictive capabilities over time.

\emph{5) MTGNN Learning Algorithm}: The learning algorithm employed by MTGNN, outlined in Algorithm \ref{alg1}, which is designed to handle large-scale graph data efficiently.  It leverages batch processing, where nodes are grouped randomly to prevent memory overflow issues.  The model learns the temporal dependencies within the network traffic data by processing batches iteratively, updating the model’s parameters based on the loss computed for each batch.

The learning algorithm incorporates the real-time feedback mechanism described above.  In each iteration, the model compares its predictions with the actual feedback and updates its parameters using the computed error.  This integration ensures that the model continuously improves its predictions by learning from real-time network data.

\begin{algorithm}
\caption{The Optimized Learning Algorithm of MTGNN with Incremental Training and Real-time Feedback}
\label{alg1}
\begin{algorithmic}[1] 
\small  
\STATE \textbf{Input:} The dataset $O$, node set $V$, the MTGNN model $f\left( \centerdot  \right)$ with $\Theta $, learning rate $\gamma $, batch size $b$, step size $s$, split size $m$, real-time feedback $y_{real}$.
\STATE set $iter=1$, $r=1$
\WHILE{non-convergence}
    \STATE sample a batch ($\chi \in {{R}^{b \times T \times N \times D}}, y_{pred} \in {{R}^{b \times T' \times N}}$) from $O$.
    \STATE random split the node set $V$ into $m$ groups, $\bigcup\nolimits_{i=1}^{m}{{{V}_{i}}=V}$.
    \IF{$iter \% s == 0$ and $r \le T'$}
        \STATE $r = r + 1$
    \ENDIF
    \FOR{$i$ in 1:$m$}
        \STATE compute $\hat{y} = f(\chi[:,:,:id({V}_{i}),:], \Theta)$.
        \STATE compute $L = loss(\hat{y}[:, :r, :], y_{pred}[:,:r, id({V}_{i})])$.
        \STATE compute the stochastic gradient of $\Theta$ according to $L$.
        \STATE update model parameters $\Theta$ using the computed gradient and the learning rate $\gamma$.
        \ENDFOR
    \STATE \textbf{Real-time Feedback Adjustment}:  
    \IF{real-time feedback $y_{real}$ is available}
        \STATE calculate the prediction error: $error = y_{real} - y_{pred}$.
        \STATE update model parameters $\Theta$ using error feedback: $\Theta = \Theta - \gamma \nabla_{\Theta} error$.
    \ENDIF
    \STATE $iter = iter + 1$.
\ENDWHILE
\end{algorithmic}
\end{algorithm}

Additionally, a curriculum learning strategy is introduced to optimize the model's performance in multi-step prediction tasks. This strategy starts with a simple prediction task and gradually increases the prediction length so that the model can learn and improve the accuracy of long-term prediction. Then, it stabilizes in a better local optimal state. 

This learning algorithm not only improves the performance of the model but also enhances its stability and generalization abilities, making it well-suited for handling diverse and complex graph data.

\emph{6) High-load Switch Identification}: To get the amount of traffic handled by each switch, we first need to associate each link with its corresponding switch.       
Once future traffic data has been predicted for each link, we match it with the switches to calculate the amount of traffic handled by each switch.       
Since a switch may be on multiple links, we must aggregate all traffic data involving that switch to get an accurate total amount of traffic.       
Then, we compare the switches' traffic to a predetermined threshold, which we set to 80\% of the switch's maximum capacity, to identify high-load switches.

\vspace{-0.5em}
\section{Network Pruning Module}
\label{section5}
After finding out the future high-load switches, the next step is generating network slices. This section presents the network pruning algorithm. The steps include subconnected graph generation, articulation point detection, and biconnected graph generation.

\begin{figure*}[htbp]
\centering
\setlength{\abovecaptionskip}{0.cm}

\includegraphics[width=\linewidth]{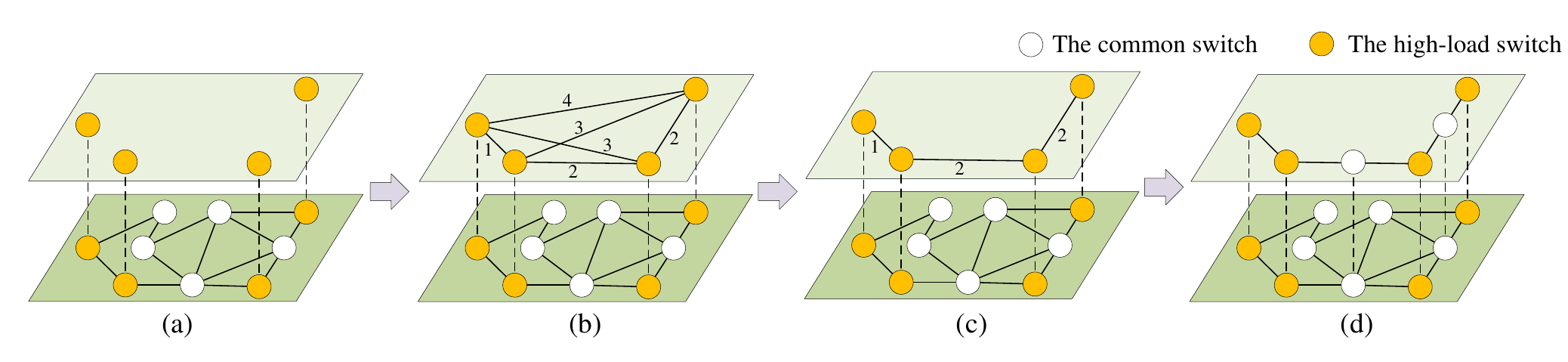}
\caption{The process of subconnected graph generation. }
\label{fig6}
\vspace{-0.1cm}
\end{figure*}
\vspace{-1em}
\subsection{Module Overview}
The network pruning module is tasked with the generation of network slicing through the network slicing algorithm.   In this module, we establish two key design objectives.

Firstly, we aim to minimize the coverage of network slices to reduce the complexity of probe path planning and maximize the utilization efficiency of network resources.

Secondly, network slices need to ensure that backup paths are available during a link/node failure. In other words, network slices should have the characteristics of the biconnected graph to enhance the reliability and fault tolerance of network telemetry.

To achieve these objectives, we design the network pruning algorithm, which realizes that backup paths are available and the coverage area is as small as possible based on covering all high-load switches.

\vspace{-0.5em}
\subsection{Subconnected Graph Generation}

The subconnected graph generation process is designed to generate a subconnected graph that can cover all high-load switches while minimizing its coverage area. At the same time, we also need to consider the connectivity of the network, ensuring that there is a valid communication path between any two target switches in the network slice.

Next, as shown in Fig. \ref{fig6}, we cover the details of each step of the subconnected graph process:

\emph{1) Create a fully connected graph with high-load switches}:
Regardless of the topology of the original network, we create a fully connected graph around high-load switches. As shown in Fig. \ref{fig6}.b, in this fully connected graph, each high-load switch is a node, and there is an edge directly connected between any two nodes.

\emph{2) Set weights for the edges of the fully connected graph}:
In the fully connected graph, each edge represents a potential path between two high-load switches. To accurately reflect the actual lengths of these paths in the original network, it is crucial to set weights for each edge. Specifically, by the shortest path algorithm such as Dijkstra or Floyd-Warshall \cite{43}, we calculate the shortest path lengths between any two high-load switches in the original network and set these lengths as weights to the corresponding edges in the fully connected graph. In Fig. \ref{fig6}.b, the number next to the edge of the fully connected graph is the weight of the edge.

\emph{3) Find a minimum spanning tree}:
After generating the weighted fully connected graph, our next objective is to find a minimum spanning tree that connects all high-load switches while minimizing the total weight of its edges. As shown in Fig. \ref{fig6}.c, we use the Kruskal algorithm to find the minimum spanning tree \cite{45}. These algorithms ensure that each edge added to the tree has the lowest weight among the currently available options, thereby resulting in a spanning tree with the lowest overall weight. 

\emph{3) Replace edges in spanning trees}:
As shown in Fig. \ref{fig6}.d, once the minimum spanning tree is determined, the next crucial step is to replace each edge of the tree with its corresponding shortest path in the original network.   This ensures that the resulting subconnected graph not only preserves the connectivity structure of the spanning tree but also accurately reflects the actual connection relationships between the nodes in the original network topology.   It's important to note that any overlapping paths must be merged during this replacement process to avoid redundancy.

\begin{figure}[h]
\centering
\setlength{\abovecaptionskip}{0.cm}

\includegraphics[width=\linewidth]{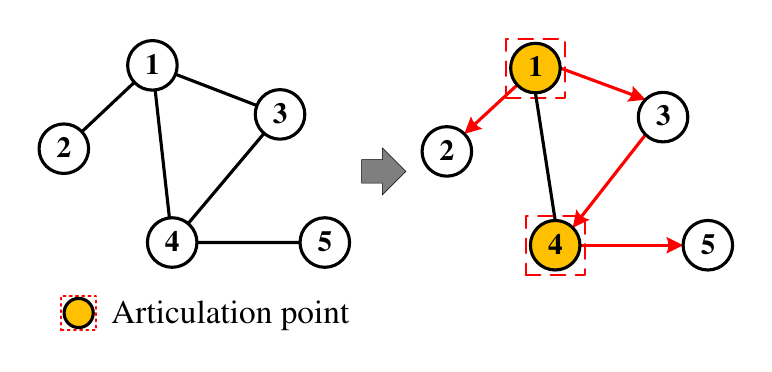}
\caption{The process of articulation point detection. }
\label{fig7}
\vspace{-0.1cm}
\end{figure}

\subsection{Articulation Point Detection}
The primary design objective of the articulation point detection process is to identify articulation points within network slices accurately. Nodes in the network are called articulation points, in which case the network slice becomes disconnected if the node is removed.
By identifying articulation nodes, we can provide an essential basis for the subsequent subnetwork optimization and fault-tolerant design.

Next, we show each step of the articulation point detection process in detail \cite{44}:

\emph{1) Step 1}:
Use the DFS algorithm to create a spanning tree $T$ called the DFS tree on the sub-connected graph.

\emph{2) Step 2}: If node $v$ of the DFS tree $T$ meets the following condition, the node $v$ is an articulation point.
\begin{itemize}  
  \item The node $v$ is the root of the DFS tree $T$, $v= {{v}_{r}}$, and node $v$ has two or more child nodes.
  \item The node $v$ is not the root of the DFS tree $T$, $v\ne {{v}_{r}}$, and node $v$ has child node $v_c$, which has no edge connected to any ancestor node of node $v$,
\end{itemize}
where $v_r$ denotes the root node of the DFS tree $T$.

As shown in Fig. \ref{fig7}, node 1 is the root node of the DFS tree, and nodes 2 and 3 are child nodes of node 1. Therefore, node 1 is an articulation point. Node 4 is not the root node, but the child node 5 of node 4 does not have any ancestor of node 4. Therefore, node 4 is also an articulation point.

\subsection{Biconnected Graph Generation}

The design of the biconnected graph generation process aims to complete the network slices into biconnected graphs \cite{28}. 
There are at least two disjoint paths between any two nodes of a biconnected graph, which means that if you remove any edge or any node in the graph, the graph is still connected. This can significantly enhance the reliability and fault tolerance of the network. Through this algorithm, we ensure that in the network slice, even if a node or link fails, the network can quickly run stably through the backup path.

The detailed steps of the biconnected graph generation process are as follows:

\emph{1) Step 1}: For each articulation point $v$, calculate the length $l(v_c, v_a)$ of the shortest path from its child node $v_c$ to each ancestor node $v_a$ of the node $v$ on the original network without using the paths contained in the DFS tree.

\emph{2) Step 2}: Select the pair of $v_c$ and $v_a$ with the smallest $l(v_c, v_a)$, and then add its path $p$ to the network slice to eliminate the articulation point.

\emph{3) Step 3}: If there are still articulation points, repeat Step 2 until there are no articulation points.

\begin{figure}
\centering
\setlength{\abovecaptionskip}{0.cm}

\includegraphics[width=\linewidth]{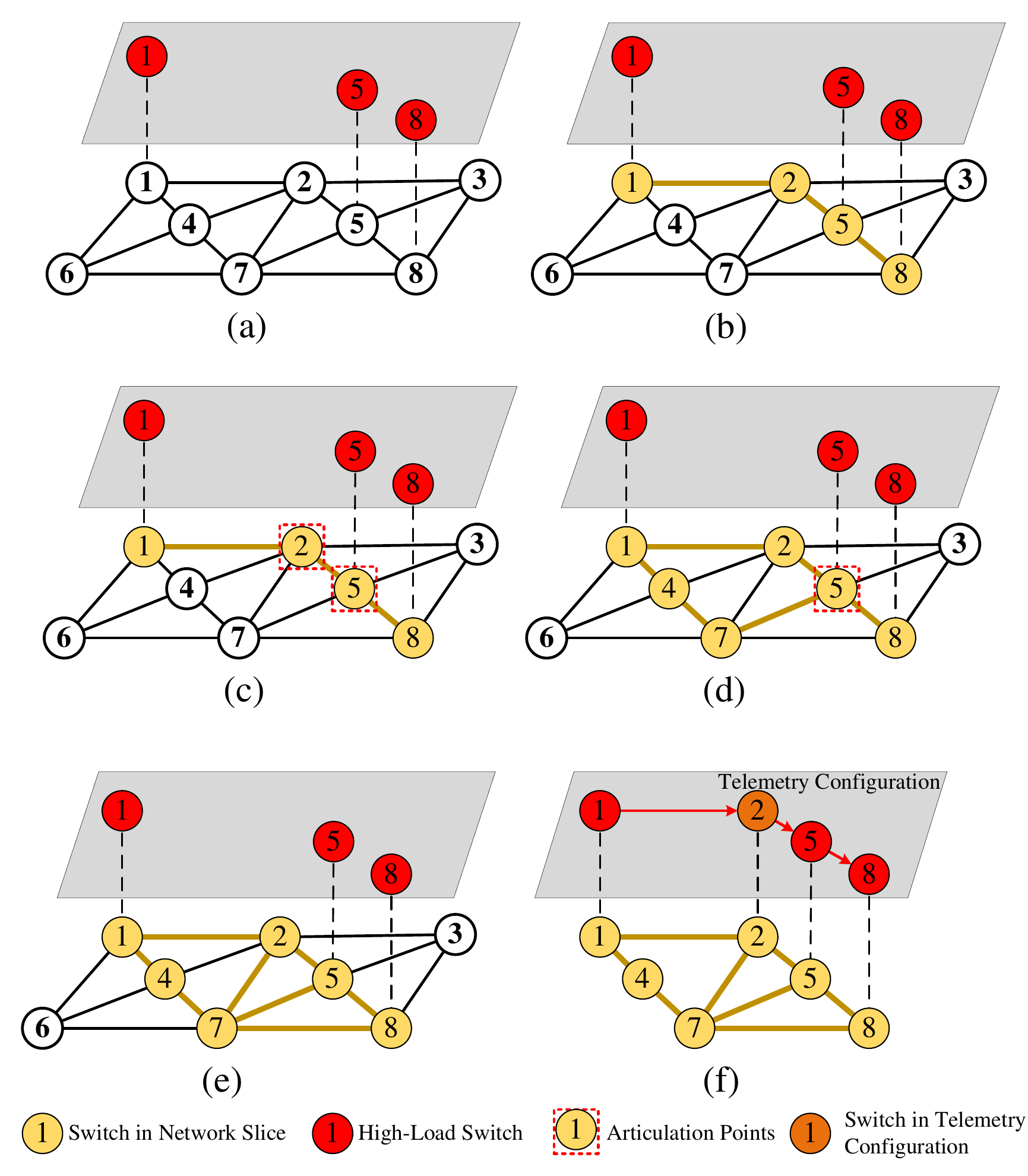}
\caption{The process of biconnected graph generation. }
\label{fig8}
\vspace{-0.1cm}
\end{figure}

Fig. \ref{fig8} shows the steps involved in the network pruning algorithm. In Fig. \ref{fig8}.a, nodes 1, 5, and 8 are identified as high-load switches. In Fig. \ref{fig8}.b, the network slice depicted represents the state achieved after executing the sub-connected graph algorithm. In Fig. \ref{fig8}.c, nodes 2 and 5 are detected as articulation points. In Fig. \ref{fig8}.d, for articulation point 5, the paths $(2,7,8)$ are selected to be added to the slice, effectively eliminating articulation point 5. In Fig. \ref{fig8}.e, for articulation point 2, the paths (1,4,7,5) are chosen for inclusion in the slice. Finally, Fig. \ref{fig8}.f represents the ultimate configuration of the network slice.
As seen from Fig. \ref{fig8}, we only need to use the necessary network slice to plan the high-frequency probe paths without using the entire underlying network.

\section{Probe Path Planning Module}
\label{section6}

In this section, we first analyze the high-frequency probe path planning problem. Then, we propose and apply a DRL model to probe path planning.

\subsection{Problem Analysis}

Considering a network topology consisting of $n$ switches, we define the network topology as an undirected physical graph, denoted by $G=\left( V, E \right)$. $V$ is the set of physical nodes represented by $V=\left\{ \left. i \right|i=1,\cdots ,n \right\}$, with $i\in V$ serving as the index for each physical node. The set of physical links is represented by $E=\left\{ \left. \left( i,j \right) \right|i,j\in V \right\}$, which comprises unordered pairs of elements from $V$. The physical link between node $i$ and node $j$ is denoted as $\left( i,j \right)$ or $\left( j, i \right)$.
Assuming that the network has ${n}'$ high-load switches in the future, we define ${{V}_{h}}$ as the set of high-load switches, ${{V}_{h}}\subseteq V$. Furthermore, the network slice created by the network slice generation module is represented as ${G}'=\left( {V}',{E}' \right)$, where ${V}'\subseteq V$ and ${E}'\subseteq E$. Since all high-load switches are covered by network slice $G'$, it follows that ${{V}_{h}}\subseteq {V}'$.

We denote the $k$-th high-frequency probe path as ${{p}_{k}}=\left[ {{v}_{k,1}},\cdots ,{{v}_{k,{{N}_{k}}}} \right]$, $k=1,2,\cdots ,K$, where ${N}_{k}$ is the number of nodes in path ${{p}_{k}}$, and ${v}_{k,{i}}$ is the $i$-th node which path ${{p}_{k}}$ passes through.
The set of switches that the high-frequency probe path $p_k$ passes through is represented as $V_k$. Based on all high-frequency probes' paths, we can represent the set $\bar{V}$ of all high-frequency probes as
\begin{IEEEeqnarray}{rCl} 
\label{form_1}
{\!}
\begin{split}
\bar{V}=\bigcup\limits_{i=1}^{K}{{{V}_{i}}}.\\
\end{split}
\end{IEEEeqnarray}
Because the high-frequency probes need to cover all high-load switches, the set $\bar{V}$ should satisfy ${{V}_{h}}\subseteq \bar{V}$.

Telemetry latency is very important for real-time network services. The latency of each link can be obtained by the network telemetry of the previous cycle. 
Define the latency function $t: E\to T$. The latency in forwarding packets from node $i$ to node $j$ is denoted by $t\left( i,j \right)$. If there is no physical link between the nodes, then $t\left( i,j \right)$ is infinite. The telemetry latency of the $k$-th probe path can be denoted as 
\begin{IEEEeqnarray}{rCl} 
\label{form_1}
{\!}
\begin{split}
{{T}_{k}}=\sum\limits_{i=1}^{{{N}_{k}}-1}{t\left( {{v}_{k,i}},{{v}_{k,i+1}} \right)},\\
\end{split}
\end{IEEEeqnarray}
where $t\left( {{v}_{k,i}},{{v}_{k,i+1}} \right)$ is the latency of $i$-th physical link on the $k$-th high-frequency probe path.
Due to the requirement of frequency consistency, the telemetry latency $T$ of the network telemetry system is the maximum probe latency, which is denoted as 
\begin{IEEEeqnarray}{rCl} 
\label{form_1}
{\!}
\begin{split}
T=\max \left\{ {{T}_{k}} \right\},k=1,2,\cdots ,K.\\
\end{split}
\end{IEEEeqnarray}
Assume that the maximum telemetry latency tolerated by the control plane is $T_{\max}$. To ensure that the control plane obtains network information in time, the telemetry latency constraint can be represented as 
\begin{IEEEeqnarray}{rCl} 
\label{form_1}
{\!}
\begin{split}
T\le {{T}_{\max }}.\\
\end{split}
\end{IEEEeqnarray}

The control overhead is crucial in active network telemetry systems, which is mainly caused by the generation and collection of probes. Therefore, the control overhead is related to the number of probes. In other words, it is determined by the number of probe paths $K$ generated by the probe path planning algorithm. The control overhead of a telemetry system can be represented as a linear function related to the number of paths, formulated as follows:
\begin{IEEEeqnarray}{rCl} 
\label{form_1}
{\!}
\begin{split}
C=a\cdot K,\\
\end{split}
\end{IEEEeqnarray}
where $a$ is the scaling factor, which quantifies the overhead per probe path.

To meet the requirement of covering all high-load switches, we focus on designing probe planning to minimize the telemetry overhead $C$. Specifically, we state the optimization problem as follows:

\begin{subequations}
\label{form_11}
\begin{equation}
\tag{\ref{form_11}}
{~~}{\underset{{{p}_{k}}}{\mathop{\min }}\,C}
\end{equation}
\vspace{-1.8em}
\begin{align}
\mathrm{s.t.}{~~}
&{{V}_{h}}\subseteq \bar{V}\subseteq {V},\label{form_11.a}\\
& T\le {{T}_{\max }}. \label{form_11.b}
\end{align}
\end{subequations}

Constraint \ref{form_11.a} ensures that the high-frequency probes cover all high-load switches. Constraint \ref{form_11.b} ensures that the telemetry latency does not exceed the maximum latency tolerated by the controller.

Problem \ref{form_11} poses a complex multipath planning challenge within a network.
DRL offers significant advantages for solving multipath planning problems.  DRL can adapt to dynamic environments and optimize path selection in complex networks.  Compared to traditional algorithms, DRL excels at handling large-scale and high-dimensional network planning issues. 

\subsection{Problem Formulation}

In this subsection, we use the DRL model to plan the probe paths. For optimization Problem \ref{form_11}, the network topology structure is obtained by the network pruning algorithm. Then, the DRL model is used to optimize the node selection on the probe paths.
Considering coverage and telemetry latency constraints, the process of probe path planning can be formulated as a constrained markov decision process (MDP), which can be defined as a tuple $\left\langle S, A, R \right\rangle $. The detailed definitions of each element are shown as follows:

\emph{1) State}: The state $s$ of DRL network at the end of slot $t$ is defined as
\begin{IEEEeqnarray}{rCl} 
\label{form_1}
{\!}
\begin{split}
{{s}}=\left\{ {{{\tilde{n}}}_{t}},{{V}_{t}} \right\},{{\tilde{n}}_{t}}\in V\bigcup{\left\{ 0 \right\}},t=0,\cdots ,T,\\
\end{split}
\end{IEEEeqnarray}
where $E_t$ represents the set of high-load switches that are not detected at time slot $t$. 
${{\tilde{n}}_{t}}\in V$ represents the current node index, and ${{\tilde{n}}_{t}}=0$ represents the creation of a new path.

\emph{2) Action}: At state $s$, the actions include adding a node to the path and creating a new path. To represent the action of creating a new path, we introduce a special variable ``0''. When the decoder selects ``0'' as the next action, it means that a new dynamic probe path is created. Thus, the action in state $s$ can be expressed as
\begin{IEEEeqnarray}{rCl} 
\label{form_1}
{\!}
\begin{split}
{{a}}=\left\{ \left. i \right|\exists \left( {{{\tilde{n}}}_{t}},i \right)\in E \right\}\bigcup{\left\{ 0 \right\}},t=0,\cdots ,T.\\
\end{split}
\end{IEEEeqnarray}
We use a greedy decoder to select the actions which can effectively improve the quality of the solution. Therefore, the action with the highest probability is chosen in each decoding step. Then, the element ${{\tilde{n}}_{t+1}}$ is defined as the action with the highest probability of the current time slot, and $E_{t+1}$ undergoes a state update based on this action.

Moreover, due to the vast number of potential actions, we implement a masking mechanism to expedite the training process. This masking scheme ensures that infeasible solutions are excluded and their log probabilities are set to negative infinity. Specifically, we apply masking to the following nodes: (i) nodes that are not in the subnetwork, (ii) nodes that lack connectivity to the current node, and (iii) nodes where telemetry requirements have already been fulfilled. 
Experiments show that the proposed masking mechanism can mask more than 90\% of the actions without affecting the quality of the solution. 
Utilizing this masking scheme not only reduces the solution space but also accelerates the discovery of a satisfactory solution. It is worth noting that when all high-load switches are covered by the probe path, all actions are masked to end the task.

\emph{3) Reward}: Considering the telemetry latency constraint, we set the negative reward function of the model according to Problem \ref{form_11}, which can be expressed as 
\begin{IEEEeqnarray}{rCl} 
\label{form_1}
{\!}
\begin{split}
r=C+\lambda \cdot flag,\\
\end{split}
\end{IEEEeqnarray}
where ${\lambda}$ is a sufficiently large constant. $flag$ is a binary variable that identifies whether the telemetry latency exceeds the preset threshold $T_{\max}$. When the telemetry latency exceeds the threshold, $flag$ is set to 1. Otherwise, $flag$ is set to 0.

\subsection{Deep Reinforcement Learning Model}

\begin{figure}
\centering
\setlength{\abovecaptionskip}{0.cm}
\includegraphics[width=\linewidth]{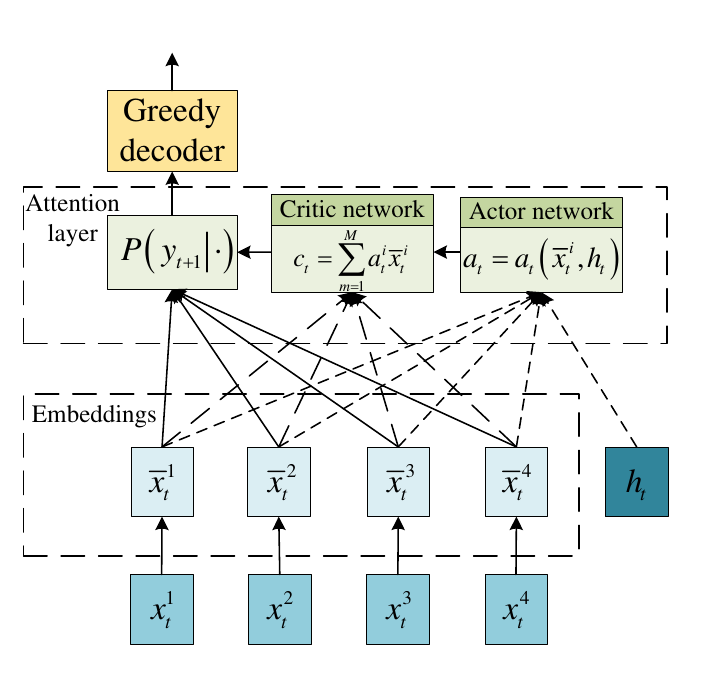}
\caption{The proposed DRL model, including an embedding layer and an attention layer. }
\label{fig9}
\vspace{-0.1cm}
\end{figure}
The DRL model we introduced is shown in Fig. 6. First, we propose a set of inputs $X\doteq \left\{ {{x}^{i}},i=0,1,\cdots,n \right\}$, where each input ${{x}^{i}}$ is a sequence of network information tuples containing information about the connection of node $i$ to other nodes and the latency information for each port.

We choose a input in ${{X}_{0}}$ as the starting point and use the pointer ${{y}_{0}}$ to identify that input. At each decoding time $t\in \left[ 0,T \right]$, we select ${{y}_{t+1}}$ from a set of available inputs ${{X}_{t}}$. This process continues until certain termination conditions are met. The sequence produced by this process can be represented as $Y=\left\{ {{y}_{t}},t=0,\cdots ,T \right\}$. We use ${{Y}_{t}}=\left\{ {{y}_{0}},\cdots ,{{y}_{t}} \right\}$ to represent the decoding sequence at time $t$. 
We need to find a random strategy ${\pi} $that is as close as possible to the optimal strategy ${{\pi}^{*}}$. Similarly to [38], we use the probability chain rule to decompose the probability $P\left(\left. y \right|{{X}_{0}} \right)$of generating sequence $Y$, as follows:
\begin{IEEEeqnarray}{rCl} 
\label{form_1}
{\!}
\begin{split}
P\left( \left. Y \right|{{X}_{0}} \right)=\prod\limits_{t=0}^{T}{P\left( \left. {{y}_{t+1}} \right|{{Y}_{t}},{{X}_{t}} \right)},\\
\end{split}
\end{IEEEeqnarray}
where $P\left( \left. {{y}_{t+1}} \right|{{Y}_{t}},{{X}_{t}} \right)$ is calculated by the attention mechanism. We denote the affine function that outputs an input size vector as $g$, and the state of the RNN decoder as $h_t$, which summarizes the information from the previous decoding step ${{y}_{0}},\cdots,{{y}_{t}}$. $P\left( \left. {{y}_{t+1}} \right|{{Y}_{t}},{{X}_{t}} \right)$ can be defined as
\begin{IEEEeqnarray}{rCl} 
\label{form_1}
{\!}
\begin{split}
P\left( \left. {{y}_{t+1}} \right|{{Y}_{t}},{{X}_{t}} \right)=\text{softmax}\left( g\left( {{h}_{t}},{{X}_{t}} \right) \right).\\
\end{split}
\end{IEEEeqnarray}
In addition, the recursive update of the problem representation  can be expressed with a state transition function $f$, as follows:
\begin{IEEEeqnarray}{rCl} 
\label{form_1}
{\!}
\begin{split}
{{X}_{t+1}}=f\left( {{y}_{t+1}},{{X}_{t}} \right).\\
\end{split}
\end{IEEEeqnarray}

The RNN encoders are highly complex due to their focus on input order, which is crucial for tasks such as text translation. However, in the probe path planning problem, we do not need to focus on the order of input node information. Therefore, as shown in Fig. \ref{fig9}, we remove the RNN encoder and perform input embedding directly using a 1-dimensional convolution layer to reduce the model complexity.

\emph{1) Attention Mechanism}: The attention layer in Fig. \ref{fig9} shows the attention mechanism of the proposed model. Similar to \cite{46}, we use a content-based attention mechanism to extract relevant information from the inputs at decoder step $i$. The variable-length alignment vector $a_t$ is used to compute this mechanism, and $\bar{x}_{t}^{i}$ represents the embedded input $x_{t}^{i}$. In addition, $h_t$ represents the memory state of the RNN cell at the decoding step $t$. The variable-length alignment vector $a_t$ determines the relevance of each input data point for the upcoming decoding step $t$, which can be expressed as
\begin{IEEEeqnarray}{rCl} 
\label{form_1}
{\!}
\begin{split}
{{a}_{t}}={{a}_{t}}\left( \bar{x}_{t}^{i},{{h}_{t}} \right)=\text{softmax}\left( {{u}_{t}} \right),\\
\end{split}
\end{IEEEeqnarray}
where $u_{t}^{i}=v_{a}^{T}\tanh \left( {{W}_{a}}\left[ \bar{x}_{t}^{i};{{h}_{t}} \right] \right)$. The symbol ``;'' denotes the concatenation of two vectors. The variables ${{v}_{a}}$ and $W_a$ are trainable variables.

We compute the conditional probability by the context vector $c_t$, and $c_t$ can be expressed as
\begin{IEEEeqnarray}{rCl} 
\label{form_1}
{\!}
\begin{split}
{{c}_{t}}=\sum\limits_{m=1}^{M}{a_{t}^{i}\bar{x}_{t}^{i}},\\
\end{split}
\end{IEEEeqnarray}
Then, using the softmax function to normalize the values, we get the following conditional probabilities:
\begin{IEEEeqnarray}{rCl} 
\label{form_1}
{\!}
\begin{split}
P\left( \left. {{y}_{t+1}} \right|{{Y}_{t}},{{X}_{t}} \right)=\text{softmax}\left( \tilde{u}_{t}^{i} \right),\\
\end{split}
\end{IEEEeqnarray}
where $\tilde{u}_{t}^{i}=v_{c}^{T}\tanh \left( {{W}_{c}}\left[ \bar{x}_{t}^{i};{{c}_{t}} \right] \right)$.  The variables ${{v}_{c}}$ and $W_c$ are trainable variables.

\begin{algorithm}
\caption{Reinforcement Learning Algorithm}
\label{alg2}
\begin{algorithmic}[1] 
\STATE \textbf{Initialization:} Initialize the actor network and critic network with random weights ${{\theta }}$ and ${{\delta }}$. 
\FOR {$i=1,2,\cdots ,epoch$}
\STATE Reset gradients. $d\theta \leftarrow 0$, $d\delta \leftarrow 0$
\STATE Sample instances  from set $\mathbf{M}$.
\FORALL {instances $\mathbf{m}=1,2,\cdots ,batch$}
\STATE $t\leftarrow0$.
\WHILE{termination condition is not reached,}
\STATE Choose the next node according to the output probabilities $P\left( \left. {{y}_{t+1}} \right|{{\mathcal Y}_{t}},{{\mathcal X}_{t}} \right)$.
\STATE Get the new state ${\mathcal X}_{t+1}$ .
\STATE $t\leftarrow t+1$.
\ENDWHILE
\STATE Compute the reward ${{R}^{\mathbf{m}}}$ based on the generated policy.
\ENDFOR
\STATE Compute $d\theta$ and $d\delta$ according to the rewards.
\STATE $d\theta \leftarrow \frac{1}{batch}\sum\limits_{\mathbf{m}=1}^{batch}{\left( {{R}^{\mathbf{m}}}-V\left( X_{0}^{\mathbf{m}};\delta  \right) \right)}{{\nabla }_{\theta }}\log P\left( \left. {{Y}^{\mathbf{m}}} \right|X_{0}^{\mathbf{m}} \right)$
\STATE $d\delta \leftarrow \frac{1}{batch}{{\sum\limits_{\mathbf{m}=1}^{batch}{{{\nabla }_{\delta }}\left( {{R}^{\mathbf{m}}}-V\left( X_{0}^{\mathbf{m}};\delta  \right) \right)}}^{2}}$
\STATE Update $\theta$ and $\delta$ according to $d\theta$ and $d\delta$.
\ENDFOR

\end{algorithmic}
\end{algorithm}

\emph{2) Training Method}:
We utilize the policy gradient method to train the network, a standard reinforcement learning approach. This method aims to optimize the policy by computing the gradient of the expected reward for the policy parameters. The policy gradient algorithm comprises two components: an actor network responsible for predicting the action probability distribution and a critic network estimating the reward for a given state. The critic network is structured with a ReLU activation layer followed by a single-output linear layer. In the Actor-Critic architecture, the critic network computes the weighted sum of the input embedding and the actor network's output, while the actor-network updates its parameters through backpropagation to enhance action selection.

Our training procedure is outlined in Algorithm \ref{alg2}, which follows a similar approach to \cite{46}. We initialize both the actor and critic networks with random weights ${{\theta }}$ and ${{\delta }}$. Then, we select instances from a set $\mathbf{M}$ of the cases for training, where the variable $batch$ represents the number of instances per training. Utilizing the output probabilities from the actor network, we generate sequences that represent policies. Once the termination condition is met, we compute the reward and update both networks accordingly. Reinforcement learning offers a suitable framework for training neural networks to tackle combinatorial optimization problems.

\vspace{-0.4em}
\section{Performance Evaluation}
\label{section7}

In this section, we first describe the simulation setup of the NTP-INT and present the benchmark schemes, followed by results and analysis.

\subsection{Simulation Setup}

In the simulation, we simulate NTP-INT in Python 3 on the platform with an Intel (R) Core (TM) i7-7700k CPU @ 4.20GHz machine equipped with 8GB RAM. The evaluation of NTP-INT includes two parts: 1) the network traffic prediction model; and 2) the probe path planning model, it should be noted that the analysis of the network pruning module will be discussed in the second part.

We generate the traffic data based on the open-source code of HPCC \cite{2}. HPCC is a simulation library that integrates RDMA based on NS-3, and it uses traffic distribution files to generate data streams with a size distribution similar to that of Alibaba's distributed storage system.
To enhance the practical value of our research findings, we also employed a dataset from a real-world network environment, known as the ``Géant" network \cite{47}, to validate the traffic prediction capabilities of various models.

We set the input sequence to 187 time slots and the output sequence to 1,4,8 and 16 time slots, respectively. As shown in Table \ref{tab1}, the model is trained with a dynamic learning rate, where $lr\left( 20 \right)$ represents the value of the learning rate when the epoch is 20. As the epoch increases, the learning rate gradually decreases, which can avoid shocks and overfitting during training. In addition, the learning rate can be adjusted according to the actual situation to improve the efficiency and performance of training. The identification threshold for high-load switches is 80\% of the highest load.

\begin{table}[htbp]  
  \centering  
  \caption{Setting of Learning Rate}  
  \label{tab1}  
  \begin{tabular}{cc}  
    \toprule 
    Epoch & Learning rate  \\  
    \midrule 
    $epoch\le 10$ & 0.0001  \\  
    $10<epoch\le 20$ & $0.0001\times {{0.95}^{epoch-10}}$  \\ 
    $20<epoch\le 30$ & $lr\left( 20 \right)\times {{0.9}^{epoch-20}}$  \\ 
    $30<epoch\le 40$ & $lr\left( 30 \right)\times {{0.9}^{epoch-30}}$  \\ 
    $40<epoch\le 50$ & $lr\left( 40 \right)\times {{0.9}^{epoch-40}}$  \\ 
    \bottomrule 
  \end{tabular}  
\end{table}  

Performance comparison benchmark algorithms are as follows.

\begin{enumerate} 
\item \emph{Graph-Wavenet:} As a predecessor to MTGNN, Graph-Wavenet replaced the convolution module with a Graph convolution module based on Wavenet.   It also uses dilatative convolution and residual connection techniques to model long data sequences effectively \cite{39}.
\item \emph{LSTNet:} LSTNet is a time series prediction model that combines CNN and RNN. The model can capture both long-term and short-term time dependencies in the data, providing more accurate predictions \cite{40}.
\item \emph{No-model:} Unlike the above two model-based traffic prediction methods, the no-model method does not rely on a specific traffic prediction model. It makes telemetry strategy directly based on current network traffic information.
\end{enumerate} 

In the simulation of the NTP-INT, we use the network traffic prediction module and Network pruning module to assist path planning.
Specifically, 64,000 instances of the network topology were created during training. Our model was trained for 20 epochs with a batch size of 1280. Both the actor network and the critic network had a learning rate of 0.0001.

The benchmark algorithms of the probe path planning algorithm are as follows.

\begin{enumerate} 
\item \emph{IntOpt:} IntOpt is an ANT system designed for NFV service chain network monitoring. It uses a stochastic greedy meta-heuristic algorithm based on simulated annealing to minimize the overhead in detection and collection \cite{8}. 
\item \emph{Depth-First-Search (DFS):} The DFS algorithm was adopted by INT-path as the probe path planning algorithm. This algorithm relies on a stack or recursive mechanism to track the order of node visits.   This approach ensures the integrity and accuracy of the probe path-planning process \cite{12}.
\item \emph{Euler Trail/Circuit:} INT-path also proposes the Euler Trail/Circuit algorithm to minimize the number of paths. The algorithm ensures that the probe path can effectively cover all network devices and improves the efficiency of network telemetry \cite{12}. 
\item \emph{NetView:} NetView presents a series of probe path planning algorithms designed for a single front-end server scenario. NetView uses the shortest path algorithm to plan the probe path after determining the target node until all target nodes are covered \cite{13}.
\item \emph{AdapINT:} AdapINT uses the DRL model directly in the basic topologies without network pruning \cite{AdapINT}.
\end{enumerate}

\subsection{Results and Analysis}

\begin{figure}
\centering
\setlength{\abovecaptionskip}{0.cm}
\includegraphics[width=\linewidth]{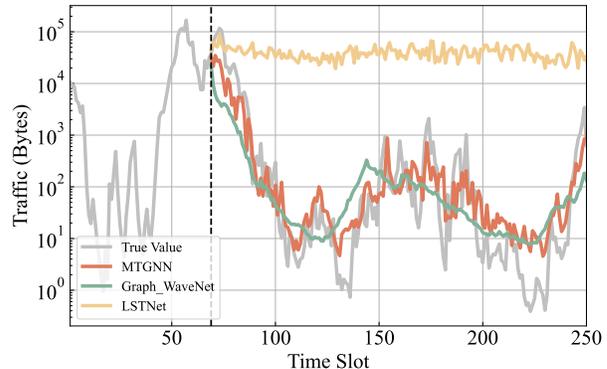}
\caption{Network traffic prediction results. }
\label{fig10}
\vspace{-0.1cm}
\end{figure}

Fig. \ref{fig10} shows the prediction results of different network traffic prediction models for a certain link, in which the predicted value is output from the 70th time slot. We can see that both MTGNN and Graph-Wavenet can capture traffic trends. LSTNet has difficulty predicting traffic accurately. This is because a large number of links bring huge input data, but the structure of LSTNet makes it challenging to deal with the relationship between multiple nodes, and the difficulty of model training is significantly increased. As a result, LSTNet's predictive power is poor.

Next, we used Mean Absolute Error (MAE) and Mean Square Error (MSE) to analyze the traffic prediction ability of each model quantitatively. Table \ref{tab2} shows the MAE and MSE values of different models with different prediction steps. We can find that with the increase in the number of prediction steps, the prediction difficulty increases, and each model's prediction ability weakens. Among them, MTGNN consistently outperforms Graph-Wavenet and LSTNet. Graph-WaveNet is superior to LSTNet. This is because both MTGNN and Graph WaveNet are GNN-based structures that very well capture the topology of the network and the hidden relationships between nodes. However, MTGNN is better at using better feature extraction modules, such as the more efficient time convolution module, graph convolution module, and graph learning layer.

\begin{table}  
\centering 
\caption{MAE and MSE of Different Traffic Prediction Models} 
\label{tab2} 
\begin{tabular}{c|cc|cc|cc}   
\hline   
\multirow{2}{*}{Step} & \multicolumn{2}{c|}{MTGNN} & \multicolumn{2}{c|}{Graph\_WaveNet} & \multicolumn{2}{c}{LSTNet} \\   
\cline{2-7} 
 & MAE & MSE & MAE & MSE & MAE & MSE \\   
\hline  

1 & 0.1519 & 0.1355 & 0.3738 & 1.1693 & 0.3108 & 1.2558 \\  
4 & 0.1853 & 0.3259 & 0.4776 & 1.7189 & 1.4726 & 8.0618 \\  
8 & 0.2805 & 0.5877 & 0.5917 & 2.5794 & 1.6772 & 9.4682 \\  
16 & 0.3693 & 0.9479 & 0.9553 & 4.9530 & 1.8803 & 11.1453 \\  
\hline  
\end{tabular}  
\end{table}  

\begin{figure}[htbp]
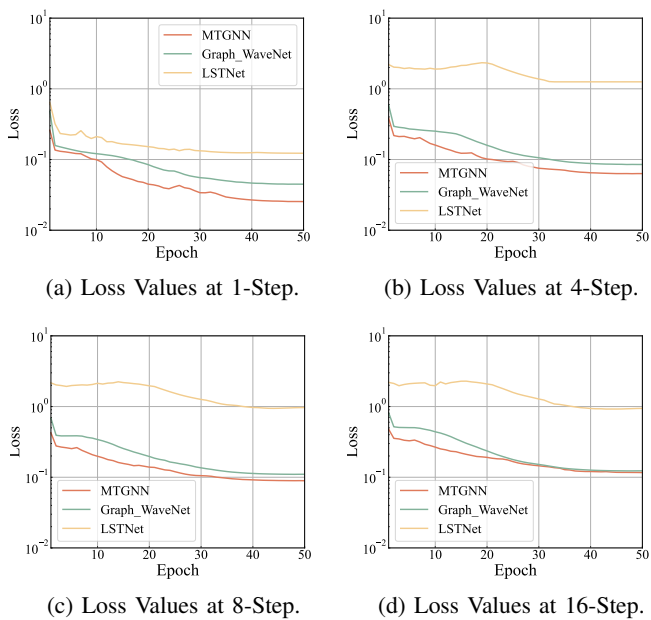
 
    \centering 
    \begin{subfigure}[b]{0.24\textwidth} 
        \includegraphics[width=\linewidth]{fig/loss\_out1.pdf} 
        \caption{Loss Values at 1-Step.} 
        \label{fig11} 
    \end{subfigure}  
    \hfill 
    \begin{subfigure}[b]{0.24\textwidth}  
        \includegraphics[width=\linewidth]{fig/loss\_out4.pdf}  
        \caption{Loss Values at 4-Step.}  
        \label{fig12}  
    \end{subfigure}  
      
    \hfill       
    \begin{subfigure}[b]{0.24\textwidth}  
        \includegraphics[width=\linewidth]{fig/loss\_out8.pdf}  
        \caption{Loss Values at 8-Step.}  
        \label{fig13}  
    \end{subfigure}  
    \hfill  
    \begin{subfigure}[b]{0.24\textwidth}  
        \includegraphics[width=\linewidth]{fig/loss\_out16.pdf}  
        \caption{Loss Values at 16-Step.}  
        \label{fig14}  
    \end{subfigure}  
      
    \caption{Loss values for different prediction steps} 
    \label{fig:main} 
\end{figure}

For the case where the prediction steps are 1, 4, 8, and 16 time slots, Fig. \ref{fig11} to Fig. \ref{fig14} respectively compare different prediction models' loss values. We can see that the loss values of all models decrease as the epoch increases. The loss values of MTGNN are lower than those of Graph\_WaveNet. The loss values of the LSTNet network decrease slowly during the training process, and there is a bottleneck that cannot be further reduced. At the same time, the prediction step also affects the training effect of the models. With the increase of the prediction step, the final loss values will become more extensive, and the training effect will become worse. This is because the amount of data the model needs to predict increases, and the training difficulty of the model increases. If the model parameters are not extended, the model's learning effect will worsen.

Table \ref{tab3} shows the precision, recall, and F1-score of traffic prediction models in identifying high-load switches. 
According to Table \ref{tab3}, we can see that MTGNN's precision, recall, and F1-score are the best. The no-model scheme has the worst performance. In other words, in the absence of network traffic prediction, it isn't easy to directly judge future network traffic based on the state of current traffic. This thoroughly explains the necessity of a network traffic prediction module.




\begin{table*}  
\centering 
\caption{Precision, recall and F1 score of different traffic prediction models} 
\label{tab3} 
\begin{tabular}{c|ccc|ccc|ccc|ccc}   
\hline   
\multirow{2}{*}{Step} & \multicolumn{3}{c|}{MTGNN} & \multicolumn{3}{c|}{Graph\_WaveNet} & \multicolumn{3}{c|}{LSTNet}& \multicolumn{3}{c}{no-model} \\   
\cline{2-13} 
 & Precision & Recall & F1 score & Precision & Recall & F1 score & Precision & Recall & F1 score & Precision & Recall & F1 score \\   
\hline  

1 & 0.8712 & 0.9319 & 0.9005 & 0.8526 & 0.8895 & 0.8707 & 0.8550 & 0.9232 & 0.8878 & 0.5048 & 0.5048 & 0.5048\\  
4 & 0.8625 & 0.9094 & 0.8853 & 0.8528 & 0.8942 & 0.8730 & 0.5607 & 0.7750 & 0.6507 & 0.5930 & 0.5935 & 0.5932\\  
8 & 0.8587 & 0.8933 & 0.8757 & 0.8512 & 0.8745 & 0.8627 & 0.4972 & 0.8593 & 0.6299 & 0.4664 & 0.4689 & 0.4677\\  
16 & 0.8511 & 0.8899 & 0.8701 & 0.7615 & 0.8981 & 0.8242 & 0.5076 & 0.8485 & 0.6352 & 0.3786 & 0.3815 & 0.3801\\  
\hline  
\end{tabular}  
\end{table*}

\begin{figure}
\centering
\setlength{\abovecaptionskip}{0.cm}

\includegraphics[width=\linewidth]{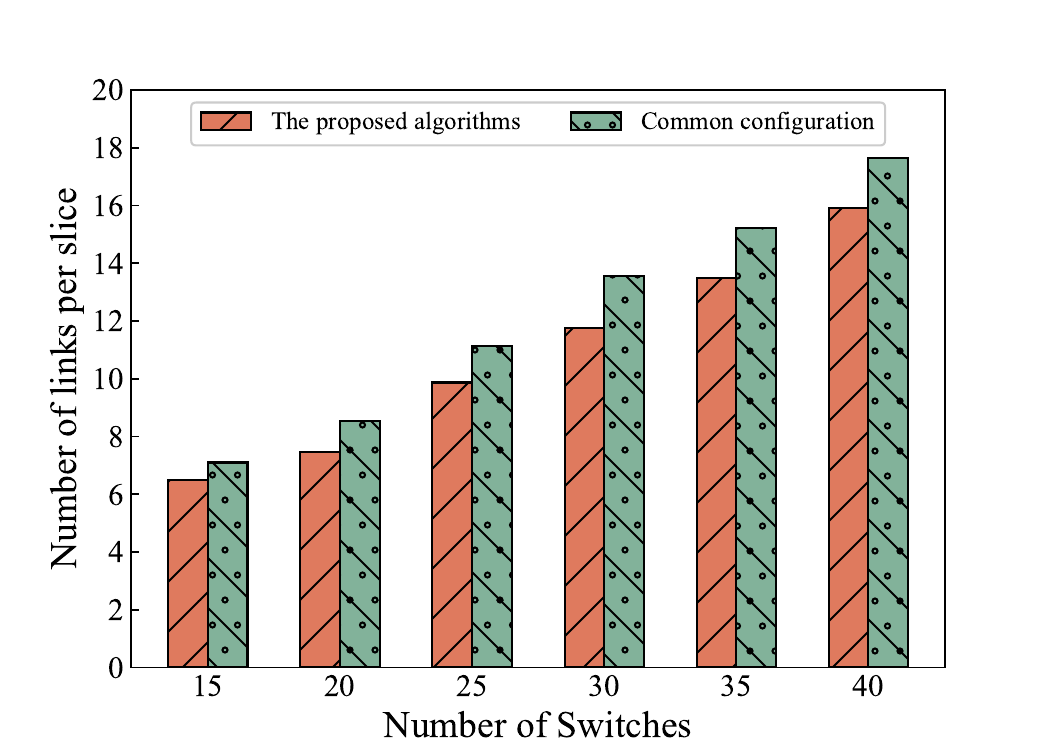}
\caption{Number of links per subnetwork. }
\label{fig15}
\vspace{-0.1cm}
\end{figure}

Fig. \ref{fig15} shows the number of links in each subnetwork. It should be noted that the common configuration scheme refers to the generation of subnets that directly use the shortest path scheme to connect high-load switches in pairs and eliminate articulation points. The network slicing algorithm can reduce the number of links in network slicing, simplifying subsequent path planning and saving computing resources effectively.

\begin{figure}
\centering
\setlength{\abovecaptionskip}{0.cm}

\includegraphics[width=\linewidth]{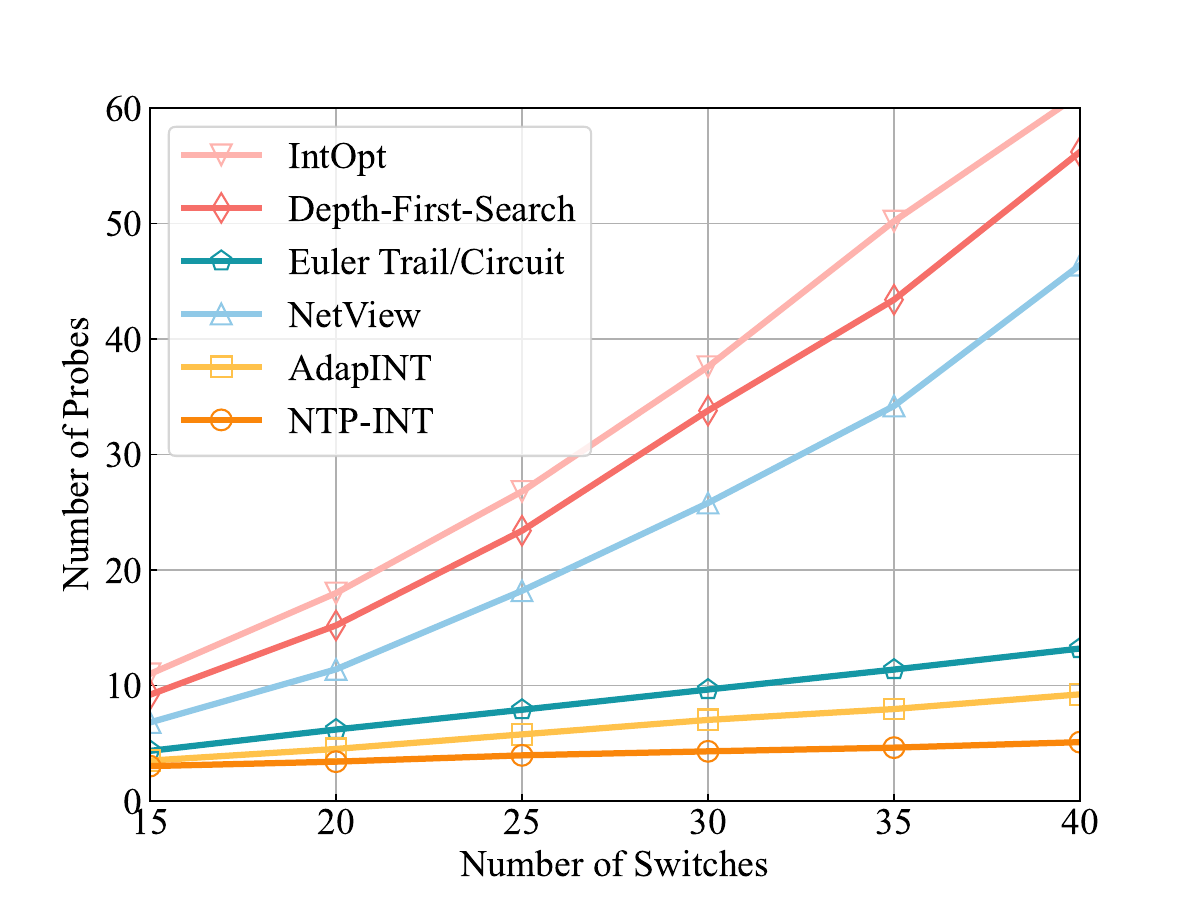}
\caption{Number of probes for different probe path planning algorithms. }
\label{fig16}
\vspace{-0.1cm}
\end{figure}

Next, we analyze the control overhead of NTP-INT. Fig. \ref{fig16} clearly shows the number of probes generated by different probe path planning algorithms directly related to the control overhead. NTP-INT is the probe path planning scheme based on network pruning technology. It is worth noting that DFS, Euler, and IntOpt are challenging to customize for high-load switches due to their inherent design principles. These three algorithms aim to cover the whole network so that the control overhead can be higher. In contrast, the NetView can carry out targeted probe path planning for high-load switches, and its control overhead is reduced compared with DFS, Euler, and IntOpt, but it is still worse than NTP-INT and AdapINT schemes. In particular, NTP-INT significantly reduces the complexity of the probe path planning problem by introducing network pruning technology, thereby improving the performance of the DRL model and achieving a smaller control overhead.

\begin{figure}
\centering
\setlength{\abovecaptionskip}{0.cm}

\includegraphics[width=\linewidth]{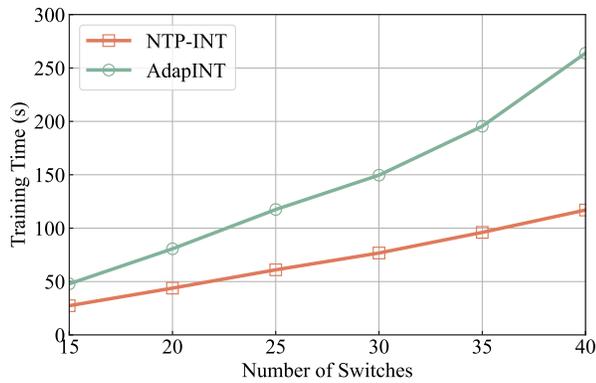}
\caption{Training time of DRL models. }
\label{fig17}
\vspace{-0.1cm}
\end{figure}

\begin{figure}
\centering
\setlength{\abovecaptionskip}{0.cm}

\includegraphics[width=\linewidth]{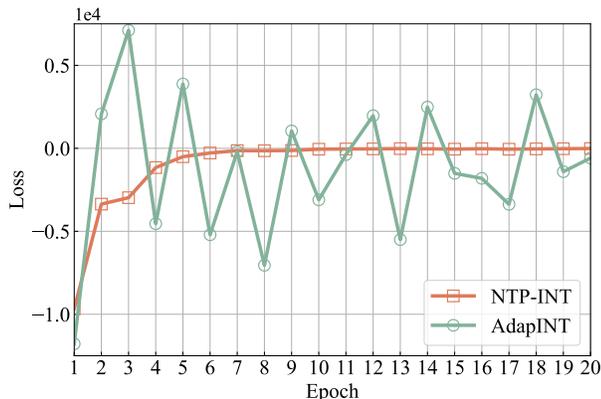}
\caption{Loss values of DRL model. }
\label{fig18}
\vspace{-0.1cm}
\end{figure}

As shown in Fig. \ref{fig17}, the network pruning module can significantly reduce the training time of DRL. Fig. \ref{fig18} shows the loss values of NTP-INT and AdapINT change with the epochs. It can be seen that the generation of the subnetwork can also significantly reduce the epochs required for the convergence of the DRL model. This greatly improves the training efficiency of the probe planning path and is essential for the model's universality.

\vspace{0em}
\section{Conclusions}
\label{section8}
\vspace{-0.2em}
In this paper, we propose NTP-INT, an intelligent network telemetry system for high-load switches, which includes the network traffic prediction module, network pruning module, and probe path planning module. By combining network traffic prediction technology and network pruning technology, the telemetry system uses a DRL-based algorithm to plan high-frequency probe paths, which can better adapt to complex dynamic network environments. The numerical results show that the system can obtain more fine-grained network information on high-load switches at the cost of small control overhead.

\bibliographystyle{ieeetr}\vspace{-0.3em}\vspace{-0.3em}
\bibliography{INT_ref1}  
\bibliographystyle{unsrt}

\end{document}